\documentclass[pra]{revtex4}
\usepackage{amssymb}
\usepackage{amsmath}
\usepackage{graphicx}

\begin{document}

\title{The creation of two-dimensional composite solitons in
spin-orbit-coupled self-attractive Bose-Einstein condensates in free space}
\author{Hidetsugu Sakaguchi and Ben Li}
\affiliation{Department of Applied Science for Electronics and Materials,
Interdisciplinary Graduate School of Engineering Sciences, Kyushu
University, Kasuga, Fukuoka 816-8580, Japan}
\author{Boris A. Malomed}
\affiliation{Department of Physical Electronics, School of Electrical Engineering,
Faculty of Engineering, Tel Aviv University, Tel Aviv 69978, Israel}

\begin{abstract}
It is commonly known that two-dimensional mean-field models of
optical and matter waves with the cubic self-attraction cannot
produce stable solitons in free space because of the occurrence of
the collapse in the same setting. By means of the numerical analysis
and variational approximation, we demonstrate that the two-component
model of the Bose-Einstein condensate with the spin-orbit Rashba
coupling and cubic attractive interactions, gives rise to
solitary-vortex complexes of two types: \textit{semi-vortices} (SVs,
with a vortex in one component and a fundamental soliton in the
other), and \textit{mixed modes} (MMs, with topological charges $0$
and $\pm 1$ mixed in both components). These two-dimensional
composite modes can be created using the trapping
harmonic-oscillator (HO) potential, but remain \emph{stable }in the
free space, if the trap is gradually removed. The SVs and MMs
realize the ground state of the system, provided that the
self-attraction in the two components is, respectively, stronger or
weaker than the cross-attraction between them. The SVs and MMs which
are not the ground states are subject to a drift instability. In the
free space (in the absence of the HO trap), modes of both types
degenerate into unstable Townes solitons when their norms attain the
respective critical values, while there is no lower existence
threshold for the stable modes. Moving free-space stable solitons
are also found in the present non-Galilean-invariant system, up to a
critical velocity. Collisions between two moving solitons lead to
their merger into a single one.
\end{abstract}

\maketitle

\section{The model}

Emulation of various effects from condensed-matter physics in atomic
Bose-Einstein condensates (BECs), where such effects may be studied in a
simpler and clearer form, has recently attracted much attention \cite%
{emulator}. A well-known example is the implementation of spin-orbit (SO)
interactions, of both the Dresselhaus \cite{Dresselhaus} and Rashba \cite%
{Rashba} types, in a binary BEC, realized experimentally \cite{Nature}, and
elaborated in detail theoretically \cite{Rashba-BEC}, as a mixture of two
different atomic states, with a linear coupling between them induced by a
specially designed laser field. In this case, the pseudo-spinor order
parameter representing standard SO effects is {mapped into} the
two-component wave functions of the mixture (a brief review of the topic was
given in Ref. \cite{Zhai}). A similar method is also used for the emulation
of gauge fields, both Abelian and non-Abelian \cite{Abelian}, in terms of
the BEC, which is a topic of a {great current }interest~too \cite{review}.

The interplay of the SO coupling, which is a linear feature of the binary
superfluid, and the {mean-field BEC\ }nonlinearity gives rise to a number of
remarkable effects, such as vortices \cite{vortex,rf:2,Fetter}, monopoles
\cite{monopole} and skyrmions \cite{skyrmions}, multi-domain patterns \cite%
{Wang}, structures induced by nonlocal interactions~\cite{Santos},
tricritical points \cite{Pitaevskii}, solitons~{\cite{AFKP,latest},
etc. In particular, recent work \cite{Fetter} elaborates the
description of two-dimensional (2D) vortices in trapping potentials,
based on the Thomas-Fermi approximation (TFA), similar to that
developed in a more general 3D setting with a spatially modulated
strength of the self-repulsive nonlinearity in Ref.
\cite{gyroscope}. }

{The use of optical lattices gives rise to additional possibilities in this
context \cite{Liu}, including the creation of gap solitons \cite{gap-sol}.}
Moreover, the fact that the system of Gross-Pitaevskii equations (GPEs),
derived for the SO-coupled binary condensate in~Ref. \cite{AFKP}, is
tantamount to the model of the co-propagation of orthogonal polarizations of
light in twisted nonlinear optical fibers~\cite{old}, establishes a link
between nonlinear {optical media and }the SO effects in binary BEC. Another
link to optics is provided by the fact that the effective pseudospin, which
underlies the concept of the SO coupling in BEC, is also a central point in
optical emulations of graphene \cite{Rechtsman}.

While, as mentioned above, diverse vortex patterns were found in SO-coupled
BEC\ with the repulsive nonlinearity, trapped in external potentials, the
aim of the present work is to construct two-dimensional (2D) vortex solitons
created in such binary condensates by \textit{attractive} nonlinear
interactions. The consideration of this setting is suggested by the
accessibility of attractive interactions, usually supported by the Feshbach
resonance \cite{Feshbach}, in many species of atomic BEC, including $^{7}$Li %
\cite{Li}, $^{85}$Rb \cite{Rb}, Na \cite{Na}, Cs \cite{Cs}, Cr \cite{Cr}, $%
^{39}$K \cite{K}, $^{172}$Yb \cite{Yb}, and $^{168}$Er \cite{Er}.

The concept of vortex solitons supported by self-attractive nonlinearities
is a well-known one since the pioneering work \cite{Minsk}, but in free
space they are usually subject to strong splitting instabilities, even if
the form of the nonlinearity (quadratic or saturable) prevents the onset of
collapse in such media \cite{Skryabin}. Another peculiarity of the vortex
solitons is that the simple TFA, which is efficient for handling vortices in
the case of the self-repulsion, as mentioned above \cite{Fetter,gyroscope},
does not apply to self-attraction.

The stabilization of vortex solitons, against the collapse and splitting
alike, is provided by trapping potentials, which may be spatially periodic
(lattice-shaped, see reviews \cite{reviews}), or harmonic-oscillator (HO)
ones, as was studied in detail for the cubic \cite{parabolic-potential,Nal}
and quadratic \cite{we} nonlinearities. Modes with the \textquotedblleft
hidden vorticity", i.e., two-components states with vorticities $+1$ and $-1$%
, can also be stabilized by the harmonic-oscillator potential \cite{Nal}.
Although periodic potentials make the settings anisotropic, destroying the
conservation of the angular momentum, the vorticity of the corresponding
localized modes can be defined in \ the usual manner, as the phase
circulation around the vortex' pivot, divided by $2\pi $.

The main result of the present work, reported below in Section 2, is
that two different families of vortex solitons, namely
\textit{semi-vortices} (SVs, with topological charges $m=0$ and $\pm
1$ in the two components) and \textit{mixed modes} (MMs, which
combine $m=0$ and $\pm 1$ in each component) are \emph{stable} in
the 2D binary system with the Rashba coupling in the \emph{free
space}, without the support of any trapping potential. Of course, in
a real experiment, any matter-wave state is created in a trapping
potential. We actually demonstrate that the semi-vortices and mixed
modes readily self-trap under the protection of the usual isotropic
HO trapping potential, and remain stable if the potential is
gradually switched off. In this connection, it is worth to mention
that the SO coupling, induced by resonant laser beams
\cite{Nature,Rashba-BEC,Zhai}, and the HO trap may be manipulated in
a completely independent fashion.

The SV and MM realize, severally, the system's ground state in the cases
when the self-attraction is stronger or weaker than its cross-attraction
counterpart [in terms of optics \cite{Agr}, these are the
self-phase-modulation (SPM)\ and cross-phase-modulation (XPM) interactions,
respectively]. Furthermore, although the Galilean invariance of the system
is broken by the SO-coupling terms, stable moving modes of both SV and MM
types exist too, up to a certain critical velocity, which is shown in
Section 3. To the best of our knowledge, this is the first example of any
species of \emph{stable} 2D solitons in a free-space system with the
attractive local cubic nonlinearity (2D solitons can be readily stabilized
by nonlocal self-attraction terms \cite{Bang}).

The system of 2D Gross-Pitaevskii equations (GPEs) for the spinor wave
function, $\boldsymbol{\phi }=(\phi _{+},\phi _{-})$, of the binary BEC with
attractive contact interactions and the SO coupling of the Rashba type, with
strength $\lambda $, which is trapped in the isotropic HO potential of
strength $\Omega ^{2}$, is written, in the scaled form, as
\begin{eqnarray}
i\frac{\partial \phi _{+}}{\partial t} &=&-\frac{1}{2}\nabla ^{2}\phi
_{+}-(|\phi _{+}|^{2}+\gamma |\phi _{-}|^{2})\phi _{+}+\lambda \left( \frac{%
\partial \phi _{-}}{\partial x}-i\frac{\partial \phi _{-}}{\partial y}%
\right) +\frac{1}{2}\Omega ^{2}\left( x^{2}+y^{2}\right) \phi _{+},  \notag
\\
i\frac{\partial \phi _{-}}{\partial t} &=&-\frac{1}{2}\nabla ^{2}\phi
_{-}-(|\phi _{-}|^{2}+\gamma |\phi _{+}|^{2})\phi _{-}-\lambda \left( \frac{%
\partial \phi _{+}}{\partial x}+i\frac{\partial \phi _{+}}{\partial y}%
\right) +\frac{1}{2}\Omega ^{2}\left( x^{2}+y^{2}\right) \phi _{-},
\label{GPE}
\end{eqnarray}%
where $\gamma $ is the strength of the XPM interactions, while the strength
of the SPM is normalized to be $1$. By means of the remaining scaling
freedom, we also set $\lambda =1$. Below, stationary solutions of Eq. (\ref%
{GPE}) are constructed in a numerical form, and, in parallel, by means of
the variational approximation (VA, which is explicitly applied to free-space
solitons, at $\Omega =0$). Solutions for moving self-trapped modes in the
free space are also obtained numerically.

To conclude the discussion of the model, it is relevant to note that
the first experimental realization of the SO coupling in the binary
BEC included equal strengths of the Rashba and Dresselhaus couplings
\cite{Nature}. However, the possibility of the experimental
implementation of the Rashba coupling proper, which is assumed in
Eq. (\ref{GPE}), was elaborated in detail too \cite{Rashba-BEC}

\section{ Solitary vortical modes}

\subsection{Semi-vortices}

First, we note that Eq. (\ref{GPE}) admits stationary solutions in the
following form:%
\begin{equation}
\phi _{+}\left( x,y,t\right) =e^{-i\mu t}f_{1}(r^{2}),~\phi _{-}\left(
x,y,t\right) =e^{-i\mu t+i\theta }rf_{2}(r^{2}),  \label{frf}
\end{equation}%
where $\mu $ is the chemical potential of the self-trapped mode, $\left(
r,\theta \right) $ are polar coordinates in the plane of $\left( x,y\right) $%
, and real functions $f_{1,2}\left( r^{2}\right) $ obey the following
equations:
\begin{gather}
\mu f_{1}+2\left[ r^{2}\frac{d^{2}f_{1}}{d\left( r^{2}\right) ^{2}}+\frac{%
df_{1}}{d\left( r^{2}\right) }\right] +\left( f_{1}^{~2}+\gamma
r^{2}f_{2}^{~2}\right) f_{1}-2\lambda \left( r^{2}f_{2}^{~\prime
}+f_{2}^{~}\right) -\frac{1}{2}\Omega ^{2}r^{2}f_{1}=0,  \notag \\
\mu f_{2}+2\left[ r^{2}\frac{d^{2}f_{2}}{d\left( r^{2}\right) ^{2}}+2\frac{%
df_{2}}{d\left( r^{2}\right) }\right] +\left( r^{2}f_{2}^{~2}+\gamma
f_{1}^{~2}\right) f_{2}+2\lambda f_{1}^{~\prime }-\frac{1}{2}\Omega
^{2}r^{2}f_{2}=0.  \label{ff}
\end{gather}%
These solutions are built as bound states of a fundamental soliton in
component $\phi _{+}$ (with zero topological charge, $m_{+}=0$), and a
solitary vortex, with $m_{-}=1$, in $\phi _{-}$. Accordingly, composite
modes of this type may be called \textit{semi-vortices} (in Ref. \cite{rf:2}%
, similar composite modes, found in a model with the repulsive nonlinearity,
were called \textquotedblleft half vortices"). The invariance of of Eq. (\ref%
{GPE}) with respect to the transformation,
\begin{equation}
\phi _{{\large \pm }}\left( r,\theta \right) \rightarrow \phi _{{\large \mp }%
}\left( r,\pi -\theta \right) ,  \label{transform}
\end{equation}%
gives rise to a semi-vortex which is a mirror image of (\ref{frf}), with the
pair of $\left( m_{+}=0,m_{-}=1\right) $ replaced by $\left(
m_{+}=-1,m_{-}=0\right) $:%
\begin{equation}
\phi _{+}\left( x,y,t\right) =-re^{-i\mu t-i\theta }f_{2}(r^{2}),~\phi
_{-}=e^{-i\mu t}f_{1}(r^{2}).  \label{mirror}
\end{equation}

The analysis of the free-space version of Eq. (\ref{ff}), with $\Omega =0$,
at $r\rightarrow \infty $ shows that the respective asymptotic form of the
solution is (written in terms of $r$, rather than $r^{2}$)%
\begin{equation}
f_{1}^{\left( \Omega =0\right) }\approx Fr^{-1/2}e^{-\sqrt{-2\mu -\lambda
^{2}}r}\cos \left( \lambda r+\delta \right) ,~f_{2}^{\left( \Omega =0\right)
}\approx -Fr^{-3/2}e^{-\sqrt{-2\mu -\lambda ^{2}}r}\sin \left( \lambda
r+\delta \right) ,  \label{asympt}
\end{equation}%
where $F$ and $\delta $ are arbitrary real constants, in terms of the
asymptotic approximation. Thus, the localized modes exist at values of the
chemical potential%
\begin{equation}
\mu <-\lambda ^{2}/2.  \label{mu}
\end{equation}%
In fact, Eq. (\ref{asympt}) gives the asymptotic form of the free-space
solitons not only for the semi-vortices, but in the general case too.

As shown below [see Fig. \ref{fig3}(c), where the semi-vortex branch is
labeled \textquotedblleft 0", which implies that $\phi _{+}$ contains solely
the zero vorticity], the free-space semi-vortices represent the ground state
of the system, in the absence of the trapping potential, at $\gamma \leq 1$.
The coexistence of the semi-vortices in the two mutually symmetric forms, (%
\ref{frf}) and (\ref{mirror}), implies the degeneracy of the ground state,
which is possible in nonlinear systems, unlike linear ones.

Previously, composite solitons built of vortical and fundamental components
were considered in a system of XPM-coupled nonlinear Schr\"{o}dinger
equations \cite{composite}. However, as said above, the usual system with
the SPM\ and XPM terms cannot produce stable solitons in free space.

\begin{figure}[b]
\begin{center}
\includegraphics[height=7.0cm]{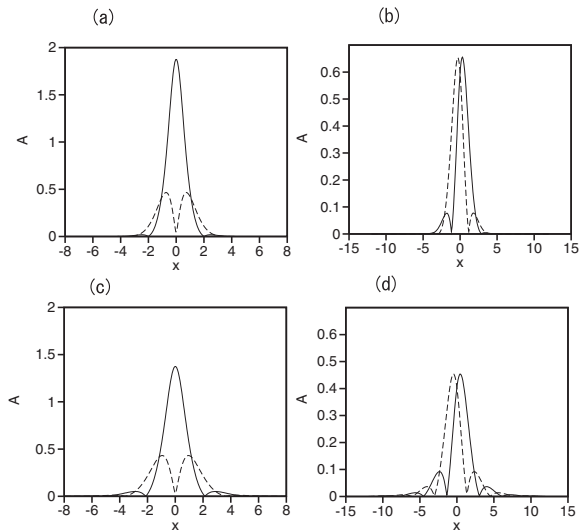}
\end{center}
\caption{(a) Cross-section profiles of the two components of a stable
semi-vortex with norm $N=5$, initiated by input (\protect\ref{00}), $|%
\protect\phi _{+}\left( x,0\right) |$ and $|\protect\phi _{-}\left(
x,0\right) |$ (the solid and dashed curves, respectively) at $\protect\gamma %
=0,\protect\lambda =1$, and $\Omega =0.5$. (b) The same for a stable mixed
mode with norm $N=2$, initiated by input (\protect\ref{mixed}), with $%
\protect\gamma =2,\protect\lambda =1$, and $\Omega =0.5$. Both solutions,
shown in (a) and (b), were generated by imaginary-time simulations. (c,d)
Shapes into which the modes from panels (a,b) relax, adiabatically following
the decrease of the trap's strength from $\Omega =0.5$ to $0$.}
\label{fig0}
\end{figure}

\begin{figure}[b]
\begin{center}
\includegraphics[height=3.5cm]{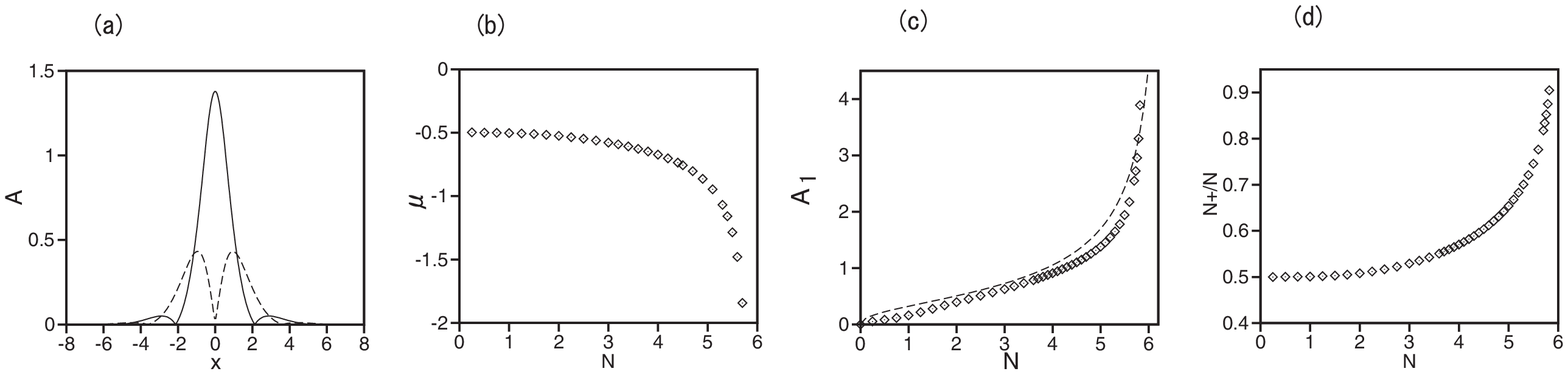}
\end{center}
\caption{Semi-vortices in the free space ($\Omega =0$). (a) The same as in
Fig. \protect\ref{fig0}(a), but for the stable semi-vortex with $\Omega =0$.
(b) Chemical potential $\protect\mu $ vs. norm $N$ for the family of
localized semi-vortices. (c) Comparison of the numerically found amplitude, $%
|\protect\phi _{+}(0,0)|$ (the chain of rhombuses), and $A_{1}$, as
predicted by the variational approximation (the dashed curve). (d) Ratio $%
N_{+}/N$ as a function of $N$, for the family of the semi-vortices.}
\label{fig1}
\end{figure}

Stable SVs were generated, as solutions to Eq. (\ref{GPE}) with $\gamma =0$
(no XPM\ nonlinearity), by means of imaginary-time simulations \cite{im-time}%
, starting from input
\begin{equation}
\phi _{+}^{(0)}=A_{1}\exp \left( -\alpha _{1}r^{2}\right) ,\;\phi
_{-}^{(0)}=A_{2}r\exp \left( i\theta -\alpha _{2}r^{2}\right) ,  \label{00}
\end{equation}%
where $A_{1,2}$ and $\alpha _{1,2}>0$ are real constants. Obviously, this
input conforms to the general ansatz (\ref{frf}) for the semi-vortices. As
shown in Figs. \ref{fig0}(a,c), to emulate an experimentally feasible
scenario, the solution was constructed in this way in the presence of the HO
potential, which was then slowly switched off, in real time. The SV
adiabatically follows the relaxation of the HO potential, remaining
dynamically stable, and eventually transforming into an SV which stays \emph{%
completely stable} in the free space, with $\Omega =0$. As said above, this
is, to the best of our knowledge, the first example of stable 2D solitons in
free space, supported by a local cubic self-attraction. In fact, as shown in
Fig. \ref{fig1}(a), the SV with the same norm,
\begin{equation}
N=\int \int (|\phi _{+}|^{2}+|\phi _{-}|^{2})dxdy\equiv N_{+}+N_{\_},
\label{N+-}
\end{equation}
can be found equally well, by means of the imaginary-time integration,
directly in the free space, setting $\Omega =0$ from the outset. It is easy
to check that not only profiles $|\phi _{+}(x,0)|$ and $|\phi _{-}(x,0)|$ of
the two field components, in the cross section of $y=0$, which are displayed
in Figs. \ref{fig0}(c) and \ref{fig1}(a), but also the entire shapes of the
SVs, which are produced, severally, by the adiabatic relaxation of the
trapping potential to $\Omega =0$ in real time, and by the direct
imaginary-time simulations in the free space, are virtually identical.

Further, Fig. \ref{fig1}(b) represents the entire family of the free-space
SVs by means of their chemical potential $\mu $, shown as a function of the
norm. Note that the $\mu (N)$ dependence satisfies the Vakhitov-Kolokolov
(VK) criterion \cite{VaKo,Berge}, $d\mu /dN<0$, which is a necessary
condition for the stability of solitary modes supported by the
self-attractive nonlinearity, although the criterion does not secure
stability of vortex solitons against splitting \cite{reviews}. We stress
that, as clearly seen in Fig. \ref{fig1}(b), there is no finite minimum
(threshold) value of $N$ necessary for the existence of the SVs in the free
space.

The wave form (\ref{00}) can be used not only as the input for the
imaginary-time simulations, but also as a variational ansatz for the
description of the free-space SVs (the VA can be extended for the model
including the HO potential, but the results are most interesting for $\Omega
=0$). Its substitution in the expression for the total energy corresponding
to Eqs. (\ref{GPE}) with $\Omega =0$,
\begin{gather}
E=\int \int \left\{ \frac{1}{2}\left( |\nabla \phi _{+}|^{2}+|\nabla \phi
_{-}|^{2}\right) -\frac{1}{2}\left( |\phi _{+}|^{4}+|\phi _{-}|^{4}\right)
-\gamma |\phi _{+}|^{2}|\phi _{-}|^{2}\right.   \notag \\
\left. +\frac{\lambda }{2}\left[ \phi _{+}^{\ast }\left( \frac{\partial \phi
_{-}}{\partial x}-i\frac{\partial \phi _{-}}{\partial y}\right) +\phi
_{-}^{\ast }\left( -\frac{\partial \phi _{+}}{\partial x}-i\frac{\partial
\phi _{+}}{\partial y}\right) \right] +\mathrm{c.c.}\right\} dxdy,  \label{E}
\end{gather}%
where $\mathrm{c.c.}$ stands for the complex conjugate, yields
\begin{equation}
E_{\mathrm{semi}}=\pi \left[ \frac{A_{1}^{2}}{2}-\frac{A_{1}^{4}}{8\alpha
_{1}}+\frac{A_{2}^{2}}{2\alpha _{2}}-\frac{A_{2}^{4}}{64\alpha _{2}^{3}}-%
\frac{\gamma A_{1}^{2}A_{2}^{2}}{4(\alpha _{1}+\alpha _{2})^{2}}+\frac{%
4\lambda A_{1}A_{2}\alpha _{1}}{(\alpha _{1}+\alpha _{2})^{2}}\right] ,
\label{Eans}
\end{equation}%
while the total norm (\ref{N+-}) of the ansatz is $N=\pi \left[
A_{1}^{2}/\left( 2\alpha _{1}\right) +A_{2}^{2}/\left( 4\alpha
_{2}^{2}\right) \right] .$ Then, values of amplitudes $A_{1},A_{2}$ and
inverse squared widths $\alpha _{1}$,$\alpha _{2}$ of the ansatz are
predicted by the minimization of $E$ with respect to the variational
parameters, $\partial E_{\mathrm{semi}}/\partial \left( A_{1,2},\alpha
_{1,2}\right) =0$. These equations\ can be easily solved numerically. Figure %
\ref{fig1}(c) displays the comparison of the so predicted amplitude $A_{1}$
and the maximum value of $|\phi _{+}|$ obtained from the
imaginary-time-generated solution at $\gamma =0$. The family of the
semi-vortices exists at $N<N_{c}\approx 5.85$, the latter value being the
well-known collapse threshold for fundamental (Townes) solitons in the free
2D space \cite{Berge}. Indeed, Fig. \ref{fig1}(d) shows, by means of the
dependence of ratio $N_{+}/N$ on $N$, that the vortical component $\phi _{-}$
vanishes at $N\rightarrow N_{c}$, hence in this limit the semi-vortex
degenerates into the usual unstable Townes soliton, which is subject to the
collapse.

In the opposite limit of $N\rightarrow 0$, the nonlinear terms in Eqs. (\ref%
{GPE}) become vanishingly small, and the ground-state solution degenerates
into a quasi-plane-wave with vanishing amplitudes, radial wavenumber $%
\lambda $ and chemical potential $\mu _{0}=-\lambda ^{2}/2$, cf. Eqs. (\ref%
{asympt}) and (\ref{mu}) \cite{rf:1}. In accordance with this expectation,
Fig. \ref{fig1}(d) shows that $N_{+}/N\rightarrow 1/2$ at $N\rightarrow 0$.
The comparison with the full numerical solutions demonstrates that Gaussian
ansatz (\ref{00}) is inaccurate for small $N$, therefore Fig. \ref{fig1}(c)
shows a large relative discrepancy between the variational and numerical
results at very small $N$.

\subsection{Mixed modes}

Another type of 2D self-trapped vortical states supported by the SO-coupled
model (\ref{GPE}) can be initiated by the following input for the
imaginary-time simulations, which may also serve as the variational ansatz:
\begin{eqnarray}
\phi _{+}^{(0)} &=&A_{1}\exp \left( -\alpha _{1}r^{2}\right) -A_{2}r\exp
\left( -i\theta -\alpha _{2}r^{2}\right) ,  \notag \\
\phi _{-}^{(0)} &=&A_{1}\exp \left( -\alpha _{1}r^{2}\right) +A_{2}r\exp
\left( i\theta -\alpha _{2}r^{2}\right) .  \label{mixed}
\end{eqnarray}%
States generated by this input are called MMs (mixed modes), as they are
built as superpositions of states with topological charges $\left(
0,-1\right) $ and $\left( 0,+1\right) $ in the two components. Unlike the
SVs, it is not possible to find an exact representation for these modes
similar to that given by Eqs. (\ref{frf}) and (\ref{ff}), but the numerical
and variational results clearly demonstrate that such states exist.
Moreover, they play the role of the ground state of the system at $\gamma
\geq 1$, see Fig. \ref{fig3}(c) below, where the mixed-mode branch is
labeled \textquotedblleft 01", as vorticities $0$ and $\pm 1$ are combined
in these modes. In accordance with the form of ansatz (\ref{mixed}), the
mixed mode is transformed into itself by symmetry reflection (\ref{transform}%
).

\begin{figure}[tbp]
\begin{center}
\includegraphics[height=3.5cm]{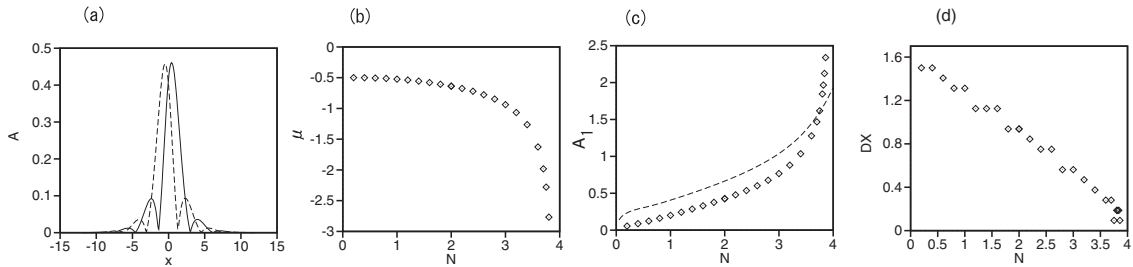}
\end{center}
\caption{Mixed modes in the free space ($\Omega =0$). (a) The same as in
Fig. \protect\ref{fig0}(b), but for the stable semi-vortex with $\Omega =0$.
(b) The chemical potential of the mixed mode vs. norm $N$ for $\protect%
\gamma =2,\protect\lambda =1$. (c) The comparison of the numerically found
central amplitude of the mixed mode, $|\protect\phi _{\pm }(0,0)|$ (the
chain of rhombuses), and its variational counterpart $A_{1}$ (the dashed
curve), vs. $N$. (d) Separation $\mathrm{DX}$ between peak positions of $|%
\protect\phi _{+}|$ and $|\protect\phi _{-}|$ vs $N$. }
\label{fig2}
\end{figure}

Similar to the SVs, Figs. \ref{fig0}(b,d) demonstrate that stable MMs can be
created under the protection of the HO potential, then adiabatically
following the decrease of the potential's strength to $\Omega =0$, while
keeping the full stability. Also similar to the SVs, Fig. \ref{fig2}(a)
demonstrates that the so transformed MMs are identical to solutions of the
same type, created directly by means of the imaginary-time integration in
free space.

The $\mu (N)$ dependence for the free-space MM family is displayed in Fig. %
\ref{fig2}(b), which shows that the VK criterion holds in this case too,
and, as well as SVs, the MMs do not require any finite threshold value of $N$
necessary for their existence either. The family exists in the interval of $%
N<N_{c}^{\prime }=2N_{c}/(1+\gamma )$, where $N_{c}$ is the above-mentioned
critical norm corresponding to the Townes solitons. Indeed, in the limit of $%
N\rightarrow N_{c}^{\prime }$ the vortical components vanish in the MM, and
it degenerates into the two-component Townes soliton, similar to the
degeneration of the SV, cf. Figs. \ref{fig1}(b) and (c). In the opposite
limit of $N\rightarrow 0$, the MM degenerates into a quasi-plane-wave with
chemical potential $-\lambda ^{2}/2$, which is again similar to the behavior
of the SV, in the same limit.

The insertion of input (\ref{mixed}), as the variational ansatz, into energy
functional (\ref{E}) yields
\begin{equation}
E_{\mathrm{mixed}}=\pi \left[ A_{1}^{2}+\frac{A_{2}^{2}}{\alpha _{2}}%
-(1+\gamma )\left( \frac{A_{1}^{4}}{4\alpha _{1}}+\frac{A_{2}^{4}}{32\alpha
_{2}^{3}}\right) -\frac{A_{1}^{2}A_{2}^{2}}{(\alpha _{1}+\alpha _{2})^{2}}+%
\frac{8\lambda A_{1}A_{2}\alpha _{1}}{(\alpha _{1}+\alpha _{2})^{2}}\right] ,
\label{Emix}
\end{equation}%
the total norm of the ansatz being $N=\pi \left[ A_{1}^{2}/\alpha
_{1}+A_{2}^{2}/\left( 2\alpha _{2}^{2}\right) \right] $. Numerical solution
of the respective energy-minimization equations, $\partial E_{\mathrm{mixed}%
}/\partial \left( A_{1,2},\alpha _{1,2}\right) =0$, produces values of the
parameters of the variational ansatz. Figure \ref{fig2}(c) compares the
absolute value of the fields at the central point, $|\phi _{+}(0,0)|$ and
its variational counterpart, $|A_{1}|$, as a function of the total norm.

As seen in Fig.~\ref{fig2}(a), peak positions of components $|\phi
_{+}\left( x,y\right) |$ and $|\phi _{-}\left( x,y\right) |$ in the MM are
separated along $x$, Fig. \ref{fig2}(d) showing the separation ($\mathrm{DX}$%
) as a function of the norm. For a small amplitude of the vortex component, $%
A_{2}$, Eq. (\ref{mixed}) yields $\mathrm{DX}\approx A_{2}/\left( \alpha
_{1}A_{1}\right) $. The separation vanishes as $N$ approaches the
aforementioned critical value $N_{c}^{\prime }$, at which the MM degenerates
into the two-component Townes solitons. This is explained by the fact that,
as said above, the vortical components of the wave functions, which cause
the shift of the peaks from the center, vanish in this limit.

\subsection{Excited states}

In addition to two types of the ground states, SVs and MMs, numerical
analysis reveals their excited varieties. First, a set of excited states can
be constructed following the general SV\ pattern given above by Eq. (\ref%
{frf}):%
\begin{equation}
\phi _{+}\left( x,y,t\right) =e^{-i\mu t+iS\theta }r^{S}f_{1}(r^{2}),~\phi
_{-}\left( x,y,t\right) =e^{-i\mu t+i\left( S+1\right) \theta
}r^{S+1}f_{2}(r^{2}),  \label{S}
\end{equation}%
with integer $S\geq 1$. The substitution of this ansatz into Eq. (\ref{GPE})
leads to a system of equations for $f_{1,2}(r^{2})$:%
\begin{gather}
\mu f_{1}+2r^{2}f_{1}^{~\prime \prime }+2\left( 1+S\right) f_{1}^{~\prime
}+r^{2S}(f_{1}^{~2}+\gamma r^{2}f_{2}^{~2})f_{1}-2\lambda \left[
r^{2}f_{2}^{~\prime }+\left( 1+S\right) f_{2}\right] -\frac{1}{2}\Omega
^{2}r^{2}f_{1}=0,  \notag \\
\mu f_{2}+2r^{2}f_{2}^{~\prime \prime }+2\left( 2+S\right) f_{2}^{~\prime
}+r^{2S}(r^{2}f_{2}^{~2}+\gamma f_{1}^{~2})f_{2}+2\lambda f_{1}^{~\prime }-%
\frac{1}{2}\Omega ^{2}r^{2}f_{2}=0.  \label{S+1}
\end{gather}%
In the case of $S=0$, Eq. (\ref{S+1}) is tantamount to Eq. (\ref{ff}) for
the SV. Of course, mirror-image counterparts of excited states (\ref{S}),
generated by transformation (\ref{transform}), exist too.

\begin{figure}[tbp]
\begin{center}
\includegraphics[height=3.5cm]{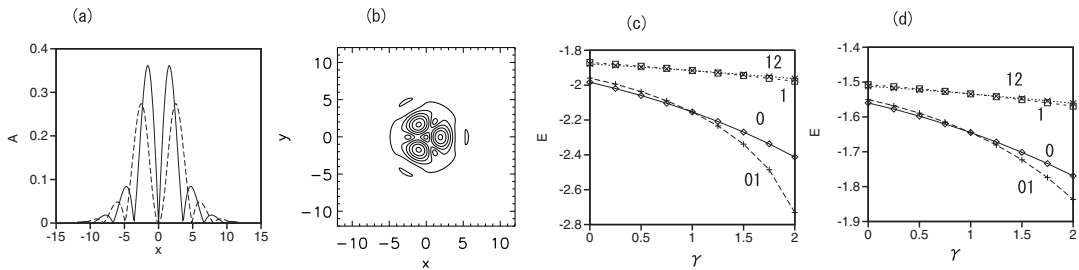}
\end{center}
\caption{(a) Cross-section profiles $|\protect\phi _{+}\left( x,0\right) |$
and $|\protect\phi _{-}\left( x,0\right) |$ (solid and dashed curves,
respectively) for the excited state in the free space, generated by input (%
\protect\ref{1}) for $N=5$ and $\protect\lambda =1,\protect\gamma =0,\Omega
=0$. (b) Contour plot of $|\protect\phi _{+}\left( x,y\right) |$ for the
excited state generated by input (\protect\ref{12}) for $N=3$ and $\protect%
\lambda =1,\protect\gamma =2,\Omega =0$. (c) Total energies for the four
free-space stationary states, viz., the semi-vortex (labeled
\textquotedblleft 0"), mixed state (\textquotedblleft 01"), and the excited
states generated by inputs (\protect\ref{1}) and (\protect\ref{12}) (labeled
\textquotedblleft 1" and \textquotedblleft 12", respectively), all taken at
a fixed value of the total norm, $N=3.7$, vs. the XPM interaction constant, $%
\protect\gamma $. (d) The same as in (c), but for $N=3$.}
\label{fig3}
\end{figure}

In the numerical form, the excited state in the free space ($\Omega =0$),
corresponding to $S=1$ in Eq. (\ref{S}), was found by means of the
imaginary-time integration starting with the following input:
\begin{equation}
\phi _{+}=A_{1}re^{i\theta }e^{-\alpha _{1}r^{2}-i\mu t},\;\phi
_{-}=A_{2}r^{2}e^{2i\theta }e^{-\alpha _{2}r^{2}-i\mu t}.  \label{1}
\end{equation}%
Figure \ref{fig3}(a) shows cross-section profiles for an example of this
excited state, obtained with $N=5$ and $\gamma =0$. In Figs. \ref{fig3}%
(c,d), the branch of these excited-state solutions is labeled
\textquotedblleft 1", as it contains vorticity $m_{+}=1$ in component $\phi
_{+}$.

Another type of excited states in the free space was generated by the input
with combined vorticities, $m_{+}=1,-2$ and $m_{-}=-1,2$, therefore it is
labeled \textquotedblleft 12" in Figs. \ref{fig3}(c,d):
\begin{eqnarray}
\phi _{+} &=&A_{1}re^{i\theta }e^{-\alpha _{1}r^{2}-i\mu
t}-A_{2}r^{2}e^{-2i\theta }e^{-\alpha _{2}r^{2}-i\mu t},  \notag \\
\phi _{-} &=&A_{1}re^{-i\theta }e^{-\alpha _{1}r^{2}-i\mu
t}+A_{2}re^{2i\theta }e^{-\alpha _{2}r^{2}-i\mu t}.  \label{12}
\end{eqnarray}%
This input can be also cast into the form of
\begin{eqnarray}
\phi _{+} &=&re^{i\theta }(A_{1}e^{-\alpha _{1}r^{2}-i\mu
t}-A_{2}r^{2}e^{-3i\theta }e^{-\alpha _{2}r^{2}-i\mu t}),  \notag \\
\phi _{-} &=&e^{-i\theta }(A_{1}e^{-\alpha _{1}r^{2}-i\mu
t}+A_{2}r^{2}e^{3i\theta }e^{-\alpha _{2}r^{2}-i\mu t}),  \label{3}
\end{eqnarray}%
which implies that it includes a vortex with topological charge $1$ set at $%
(x,y)=(0,0)$, and three vortices with charges $-1$ surrounding the origin.
Figure \ref{fig3}(b) corroborates this interpretation by means of a contour
plot of $|\phi _{+}\left( x,y\right) |$, which features three peaks and
three holes around the origin. The holes are pivots of the three
above-mentioned vortices with charges $-1$. The pattern is symmetric with
respect to rotation by angle $2\pi /3$. The respective contour map of $|\phi
_{-}\left( x,y\right) |$ (not shown here)\ is a mirror image of $\left\vert
\phi _{+}\left( x,y\right) \right\vert $, generated by transformation (\ref%
{transform}). This solution resembles a lattice state found in Ref. \cite%
{rf:3} for the SO-coupled BEC\ with the self-repulsive interactions, trapped
in the HO potential.

\subsection{The identification of the ground state, and analysis of the
stability of the vortical modes in the free space}

The four types of the vorticity-carrying self-trapped modes, generated by
inputs (\ref{frf}), (\ref{mixed}), (\ref{1}), and (\ref{12}), respectively,
in the free space ($\Omega =0$), can be produced by the imaginary-time
integration of Eq. (\ref{GPE}) for any value of the XPM-interaction
constant, $\gamma $ [in addition to the two latter modes, excited states of
still higher orders can be found too -- e.g., those given by Eq. (\ref{S})
with $S>1$-- but they all are strongly unstable]. To identify the system's
ground state, the total energies of the four species of the vortical modes,
calculated as per Eq. (\ref{E}), and denoted as $E_{0}$ (for the
semi-vortices), $E_{01}$ (for the mixed mode), and $E_{1}$, $E_{12}$ for the
excited states (\ref{1}) and (\ref{12}), respectively, are displayed vs. $%
\gamma $ in Figs. \ref{fig3}(c,d), for two fixed values of the total norm, $%
N=3.7$ and $N=3$. It is found that the energies satisfy relations $%
E_{0}<E_{01}<E_{12}<E_{1}$ at $\gamma <1$ and $E_{01}<E_{0}<E_{1}<E_{12}$ at
$\gamma >1$. It is seen that the semi-vortex and mixed state realize the
ground state at $\gamma <1$ and $\gamma >1$, respectively, while the states
labeled \textquotedblleft 1" and \textquotedblleft 12" are indeed excited
states, separated by a wide energy gap from the competing ground-state
modes. The fact that the switch of the ground states occurs at $\gamma =1$
is not surprising, as it corresponds to the Manakov's nonlinearity, with
equal XPM and SPM coefficients, which is known to feature various
degeneracies in nonlinear systems, see, e.g., Ref. \cite{Manakov}. It is
relevant to mention that the value of $\gamma $, which is the ratio of the
strengths of the XPM and SPM nonlinearity, may be readily altered by means
of the Feshbach resonance \cite{Feshbach}, hence the type of the ground
state may be controlled by means of this technique.

The stability of the four species of 2D\ self-trapped modes constructed
above was studied by means of systematic numerical simulations of their
perturbed evolution in the framework of Eq. (\ref{GPE}). The results are
reported here for the generic case, represented by two values of the XPM
coefficient, $\gamma =0$ and $2$, and a fixed norm, $N=3.7$. The first
result is that the SV, which is the ground state at $\gamma =0$, and the MM,
which plays the same role at $\gamma =2$, are stable against perturbations
(not shown here in detail, as their stability manifests itself in an obvious
way).

\begin{figure}[tbp]
\begin{center}
\includegraphics[height=4.cm]{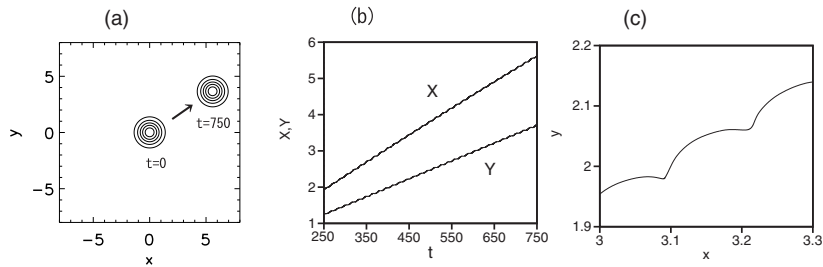}
\end{center}
\caption{(a) The evolution of the semi-vortex with $N=3.7$, initiated by
input (\protect\ref{00}) in the free space ($\Omega =0$), at $\protect\gamma %
=2$, when the semi-vortex is \emph{not} the ground state, according to Fig.
\protect\ref{fig3}(c). The contour plots of $|\protect\phi _{+}\left(
x,y\right) |$ at $t=0$ and $t=750$ are shown. (b) The time dependence of the
coordinates of the peak position, $(X,Y)$, of $|\protect\phi _{+}\left(
x,y\right) |$. (c) A zoom of a segment of the trajectory $(X(t),Y(t))$,
which demonstrates a small oscillatory component of the motion.}
\label{fig4}
\end{figure}

Next, it is interesting to test the stability of the same two species in the
cases when they are \emph{not} ground states, i.e., the SV at $\gamma =2$,
and the MM at $\gamma =0$. In the former case, we observe in Fig. \ref{fig4}%
(a) that the SV profile keeps the initial shape from $t=0$ till $t=750$,
which exceeds $100$ diffraction times for the present mode. However, an
instability manifests itself in spontaneous motion of the soliton with a
nearly constant velocity, as seen in Fig. \ref{fig4}(b), where coordinates
of the peak position of the $\phi _{+}$ component, $(X,Y)$, are shown as a
function of time. On top of the mean velocity, the peak features oscillatory
motion with a small amplitude, see Fig. \ref{fig4}(c).

\begin{figure}[t]
\begin{center}
\includegraphics[height=3.5cm]{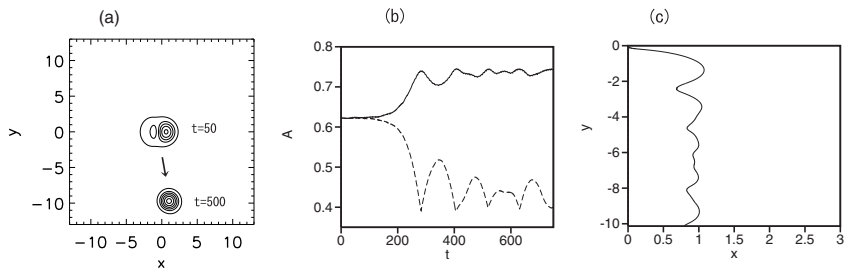}
\end{center}
\caption{(a) Contour plots of $|\protect\phi _{+}\left( x,y\right) |$ at $%
t=50$ and $t=500$ for the mixed mode generated by input (\protect\ref{mixed}%
) with $N=3.7$ in the free space ($\Omega =0$), at $\protect\gamma =0$, when
this mode is not a ground state. (b) The evolution of amplitudes of the $%
\protect\phi _{+}$ and $\protect\phi _{-}$ components (solid and dashed
curves, respectively) for $0<t<750$. (c) The trajectory of the peak position
of $|\protect\phi _{+}\left( x,y\right) |$ for $0<t<500$.}
\label{fig5}
\end{figure}

The evolution of the mixed mode at $\gamma =0$, when it is not the ground
state either, is shown in Fig. \ref{fig5}. Panels (a), pertaining to $t=50$
and $t=500$ (the latter can be estimated to be $\simeq 70$ diffraction times
of the present mode), demonstrate that this state is unstable, starting
spontaneous motion and losing the original symmetry between $|\phi _{+}|$
and $|\phi _{-}|$ with respect to to transformation (\ref{transform}). By $%
t=500$, the mixed mode rearranges into a state close to a semi-vortex.
Further, Fig. \ref{fig5}(b) shows the time evolution of amplitudes of the $%
\phi _{+}$ and $\phi _{-}$ components (the solid and dashed curves,
respectively) for $0<t<750$. Breaking the original symmetry, the amplitude
of $|\phi _{+}|$ increases toward the value of the amplitude of the $|\phi
_{+}|$ component of the semi-vortex state for the same $N=3.7$, while the
amplitude of $|\phi _{-}|$ falls to become nearly equal to the amplitude of
the $|\phi _{-}|$ component of the same SV. Figure \ref{fig5}(c) shows a
trajectory of the peak position of the $|\phi _{+}|$ component. The
localized state moves spontaneously, featuring oscillations in the $x$
direction, while the average velocity in the $y$-direction is $v_{y}=-0.0175$%
.

\begin{figure}[tbp]
\begin{center}
\includegraphics[height=3.5cm]{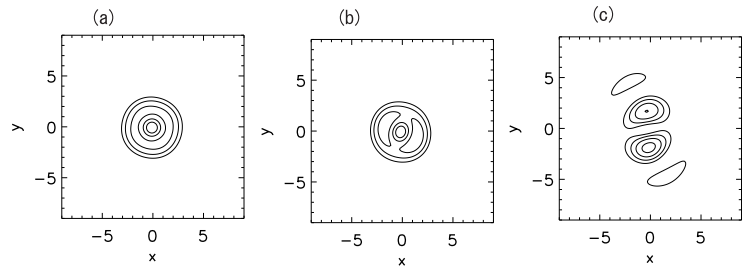}
\end{center}
\caption{Contour plot of $|\protect\phi _{+}\left( x,y\right) |$ of the
excited vortical state, generated by input (\protect\ref{1}) with $N=3.7$ in
the free space ($\Omega =0$), at $\protect\gamma =0$: (a) $t=500$, (b) $t=950
$, (c) $t=1200$. At $\protect\gamma =2$, the instability of this excited
state is essentially stronger.}
\label{fig6}
\end{figure}

The excited states generated by inputs (\ref{1}) and (\ref{12}) are unstable
both at $\gamma =0$ and $g=2$. In particular, the ring structure of the
former vortical state splits in the course of the evolution, as shown in
Fig. \ref{fig6} for $\gamma =0$. Similar splitting is observed much earlier
(already at $t=100$) for $\gamma =2$.

\begin{figure}[tbp]
\begin{center}
\includegraphics[height=3.5cm]{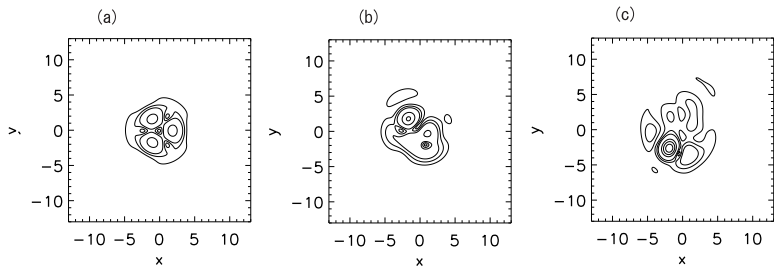}
\end{center}
\caption{Contour plot of $|\protect\phi _{+}\left( x,y\right) |$ at $t=50$
(a), $t=150$ (b), and $t=250$ (c) for the excited state  with $N=3.7$,
generated by input (\protect\ref{12}) in the free space ($\Omega =0$), at $%
\protect\gamma =2$. At $\protect\gamma =0$ this mode is unstable too.}
\label{fig7}
\end{figure}

Figure \ref{fig7} illustrates the evolution of the excited state obtained
from input (\ref{12}) at $\gamma =2$. The original vortex complex is broken
by the instability, evolving into an apparently chaotic pattern. This
instability develops quickly, as $t=250$ [Fig. \ref{fig7}(c)] corresponds to
$\lesssim 5$ diffraction times of the original structure.

Additional analysis (not shown here in detail) demonstrates that, in the
presence of the HO trapping potential, the SV and MM remain, severally,
stable at $\gamma <1$ and $\gamma >1$. On the other hand, although the
instability of the same modes in the opposite cases, i.e., $\gamma >1$ for
the SV, and $\gamma <1$ for the MM, manifests itself, in the free space, by
spontaneous drift (see Figs. \ref{fig4} and \ref{fig5}, respectively),
additional simulations demonstrate that the confining HO potential, unless
it is very tight, does not suppress the drift instability. Thus, the results
obtained in the free space adequately represent the situation in the
presence of the HO potential too.

\section{Mobility and collisions of vortex modes in the free space}

The numerical results displayed in Figs.~\ref{fig4} and \ref{fig5} suggest
that propagating modes may exist in the free space. Localized states which
move steadily at velocity $\mathbf{v}=(v_{x},v_{y})$ can be looked for in
the form of $\phi _{+}=\phi _{+}(x-v_{x}t,y-v_{y}t,t)$ and $\phi _{-}=\phi
_{-}(x-v_{x}t,y-v_{y}t,t)$. The substitution of this into Eq. (\ref{GPE})
with $\Omega =0$ leads to the equations written in the moving reference
frame,
\begin{eqnarray}
i\frac{\partial \phi _{+}}{\partial t}-i\left( \mathbf{v}\cdot \nabla
\right) \phi _{+} &=&-\frac{1}{2}\nabla ^{2}\phi _{+}-(|\phi
_{+}|^{2}+\gamma |\phi _{-}|^{2})\phi _{+}^{\prime }+\lambda \left( \frac{%
\partial \phi _{-}}{\partial x}-i\frac{\partial \phi _{-}}{\partial y}%
\right) ,  \notag \\
i\frac{\partial \phi _{-}^{\prime }}{\partial t}-i\left( \mathbf{v}\cdot
\nabla \right) \phi _{-} &=&-\frac{1}{2}\nabla ^{2}\phi _{-}-(|\phi
_{-}|^{2}+\gamma |\phi _{+}|^{2})\phi _{-}+\lambda \left( -\frac{\partial
\phi _{+}}{\partial x}-i\frac{\partial \phi _{+}}{\partial y}\right) ,
\label{moving}
\end{eqnarray}%
where $x$ and $y$ actually stand for $x-v_{x}t$ and $y-v_{y}t$. Note that
Eq. (\ref{GPE}) has no Galilean invariance, hence steadily propagating
solutions cannot by generated by a straightforward transformation, such as
\begin{equation}
\phi _{{\large \pm }}\left( \mathbf{r}\right) \equiv \tilde{\phi}_{{\large %
\pm }}(\mathbf{r})\exp \left( i\mathbf{v}\cdot \mathbf{r}-\frac{i}{2}%
v^{2}t\right) .  \label{Galileo}
\end{equation}%
In particular, in the case of $v_{x}=0$ (in this case, solutions for moving
modes exist, as shown below), the quasi-Galilean transformation (\ref%
{Galileo}) casts Eq. (\ref{moving}) into a form which differs from
underlying equations (\ref{GPE}) by the presence of terms causing for linear
mixing of the two components:
\begin{eqnarray}
i\frac{\partial \phi _{+}}{\partial t} &=&-\frac{1}{2}\nabla ^{2}\phi
_{+}-(|\phi _{+}|^{2}+\gamma |\phi _{-}|^{2})\phi _{+}^{\prime }+\lambda
\left( \frac{\partial \phi _{-}}{\partial x}-i\frac{\partial \phi _{-}}{%
\partial y}\right) +\lambda v_{y}\phi _{-},  \notag \\
i\frac{\partial \phi _{-}^{\prime }}{\partial t} &=&-\frac{1}{2}\nabla
^{2}\phi _{-}-(|\phi _{-}|^{2}+\gamma |\phi _{+}|^{2})\phi _{-}+\lambda
\left( -\frac{\partial \phi _{+}}{\partial x}-i\frac{\partial \phi _{+}}{%
\partial y}\right) +\lambda v_{y}\phi _{+}.  \label{mix}
\end{eqnarray}%
The same linear mixing can be imposed, in diverse 1D \cite{radio} and 2D
\cite{radio-2D} settings, by a GHz wave coupling the two underlying atomic
states, i.e., the linear mixing by itself represents a physically relevant
addition to the basic model. A straightforward impact of the addition of the
mixing terms in Eq. (\ref{mix}) is a shift of the edge of the \textit{%
semi-infinite gap} (\ref{mu}) in which solitons may exist, from $\mu
=-\lambda ^{2}/2$ to $\mu <-\left( \lambda ^{2}/2+\left\vert \lambda
v_{y}\right\vert \right) $.

\begin{figure}[tbp]
\begin{center}
\includegraphics[height=3.5cm]{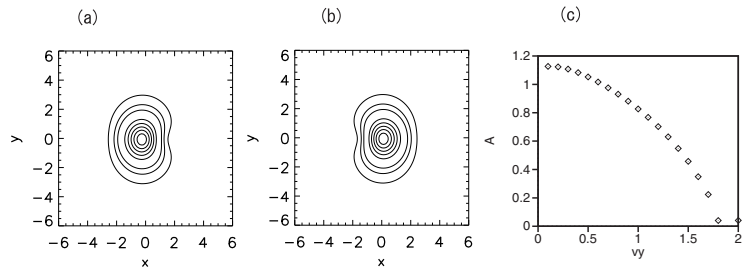}
\end{center}
\caption{Contour plots of $|\protect\phi _{+}\left( x,y\right) |$ (a) and $|%
\protect\phi _{-}\left( x,y\right) |$ of the stable mixed mode with norm $%
N=3.1$, moving at velocity $v_{y}=0.5$ (while $v_{x}=0$) in the free space ($%
\Omega =0$), for $\protect\gamma =2,\protect\lambda =1$. (c) The amplitude
of the moving mixed mode as a function of $v_{y}$, also for $N=3.1$.}
\label{fig8}
\end{figure}

Coming back to equations (\ref{moving}) written in the moving reference
frame, stationary solutions to these equations can be obtained, as well as
in the case of underlying equations (\ref{GPE}), by means of the
imaginary-time evolution method for $v_{y}\neq 0$, but the procedure
produces results solely for $v_{x}=0$ (this situation is possible, as the
present 2D system is not isotropic). In particular, at $\gamma =2$, when the
quiescent MM is the ground state, its moving version, which is displayed in
Figs. \ref{fig8}(a,b) for $N=3.1$ and $v_{y}=0.5$, exists and is stable too.
As well as its quiescent counterpart, this mode features the mirror symmetry
between the profiles of $|\phi _{+}\left( x,y\right) |$ and $|\phi
_{-}\left( x,y\right) |$. Figure \ref{fig8}(c) shows the amplitude of the
moving MM, $A=\sqrt{|\phi _{+}(x=0,y=0)|^{2}+|\phi _{-}(x=0,y=0)|^{2}}$, as
a function of $v_{y}$. The amplitude monotonously decreases with the growth
of the velocity, and the mode vanishes at $v_{y}=\left( v_{y}\right) _{\max
}^{(\mathrm{mixed})}\approx 1.8$.

The availability of the stably moving mixed modes suggests to consider
collisions between them. In particular, we have performed simulations of Eq.
(\ref{GPE}) for the head-on collision between two solitons displayed in
Figs. \ref{fig8}(a,b), moving at velocities $v_{y}=\pm 0.5$. Figures \ref%
{fig9}(a), (b), and (c) display snapshot patterns of $\sqrt{|\phi
_{+}(x,y)|^{2}+|\phi _{-}(x,y)|^{2}}$ at $t=4$, $t=24$, and $t=64$,
respectively. The collision results in fusion of the two solitons into a
single state, of the same MM type, which is spontaneously drifting along
direction $x$ with velocity $v_{x}\approx 0.14$. The drift may be understood
as a manifestation of spontaneous symmetry breaking caused by the collision.
Strictly speaking, this state, produced by the fusion, is not a steadily
moving one, because, as mentioned above, the imaginary-time integration of
Eq. (\ref{moving}) does not produce MM solitons moving in the direction of $x
$. The fusion is accompanied by emission of small-amplitude radiation waves.
\begin{figure}[tbp]
\begin{center}
\includegraphics[height=3.5cm]{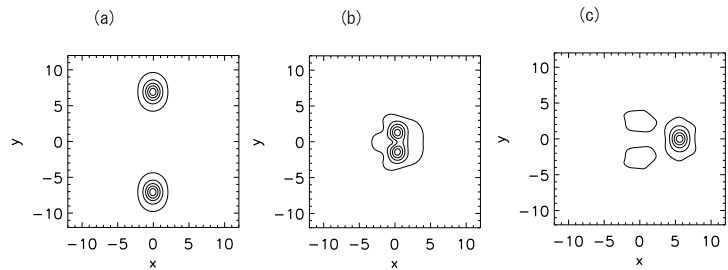}
\end{center}
\caption{The head-on collision between two mixed-mode solitons with norms $%
N=3.1$ moving, in the free space ($\Omega =0$), at velocities $v_{y}=\pm 0.5$
[the same as those shown in Fig. \protect\ref{fig8}(a,b)]. Contour plots of $%
\protect\sqrt{|\protect\phi _{+}^{\prime }\left( x,y\right) |^{2}+|\protect%
\phi _{-}\left( x,y\right) |^{2}}$ are displayed at $t=4$ (a), $t=24$ (b),
and $t=64$ (c).}
\label{fig9}
\end{figure}

\begin{figure}[tbp]
\begin{center}
\includegraphics[height=3.5cm]{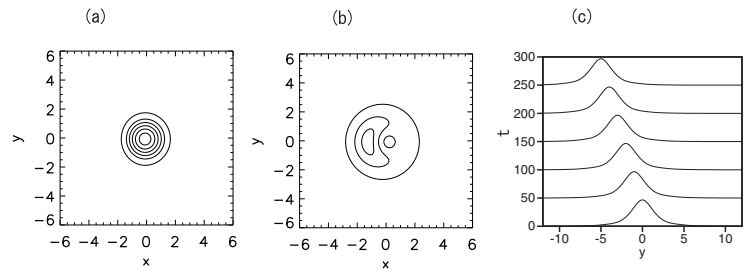}
\end{center}
\caption{Contour plots of $|\protect\phi _{+}\left( x,y\right) |$ (a) and $|%
\protect\phi _{-}\left( x,y\right) |$ for the stable semi-vortex with norm $%
N=3.7$, which moves, in the free space ($\Omega =0$), at velocity $%
v_{y}=-0.02$. (c) The evolution of $\protect\sqrt{|\protect\phi _{+}\left(
0,y\right) |^{2}+|\protect\phi _{-}\left( 0,y\right) |^{2}}$ in the cross
section drawn through $x=0$.}
\label{fig10}
\end{figure}

At $\gamma =0$, the SV, which, as shown above, is the ground state in the
class of quiescent modes in this case, can also move stably, but only in a
small interval of velocities,
\begin{equation}
|v_{y}|\leq \left( v_{y}\right) _{\max }^{(\mathrm{semi})}\approx 0.03
\label{max2}
\end{equation}%
[cf. a much larger velocity found above for the moving MM at $\gamma =2$].
Figures \ref{fig10}(a) and (b) show the profiles of $|\phi _{+}\left(
x,y\right) |$ and $|\phi _{-}\left( x,y\right) |$ for the SV with norm $N=3.7
$, moving at velocity $v_{y}=-0.02$. In addition, Fig. \ref{fig10}(c)
displays the evolution of $\sqrt{|\phi _{+}|^{2}+|\phi _{-}|^{2}}$ in the
cross section of $x=0$, produced by direct simulations of Eq.~(\ref{GPE}),
starting from the initial conditions corresponding to Figs.~\ref{fig10}%
(a,b). The localized solution is stably moving at velocity $v_{y}=-0.02$. In
fact, this moving state is similar to the one generated by the spontaneous
onset of motion of the unstable quiescent MM at $\gamma =0$ and the same
norm, which rearranges into a state close to the SV, as shown above in Fig. %
\ref{fig5}.

At $\left\vert v_{y}\right\vert >0.03$ [see Eq. (\ref{max2})], the solution
to Eq. (\ref{moving}) produced by means of the imaginary-time-propagation
method converges not to an SV, but rather to an MM state, which turns to be
stable in real-time simulations. Thus, the moving SVs are rather fragile
objects, while the MMs are, on the contrary, very robust ones in the state
of motion.

\section{Conclusions}

The objective of this work was to construct several types of self-trapped
vortex-soliton complexes in the 2D model of the binary BEC,\ with the SO
coupling of the Rashba type between the two components and attractive
intrinsic nonlinearity. The most essential finding is that, on the contrary
to the commonly known instability of 2D free-space solitons and vortices in
previously studied models with the attractive cubic terms, two species of
\emph{stable} modes have been found here in the numerical form and by means
of the VA (variational approximation), \textit{viz}., the SVs
(semi-vortices) and MMs (mixed modes), which represent the system's ground
state when the SPM attraction is, respectively, stronger or weaker than its
XPM counterpart. An experimentally relevant scenario was elaborated, which
assumes that the SVs and MMs are originally created under the protection of
the HO trapping potential, which is subsequently switched off, slowly
enough. The stable modes readily follow the lifting of the trap,
adiabatically carrying over into the stable self-trapped states which exist
in the free space.

The general structure of the SVs and similar higher-order excited states was
found in the exact form, see Eqs. (\ref{frf}), (\ref{ff}) and (\ref{S}), (%
\ref{S+1}). In the free space, the SVs and MMs turn into the commonly known
unstable Townes solitons when norms of these states approach the
corresponding limit values, while there is no lower threshold value of the
norm necessary for their existence. Moving stable modes have been found too,
up to the respective critical velocities, in spite the lack of the Galilean
invariance in the system. Collisions between moving mixed modes lead to
their fusion.

It should be relevant to extend the analysis by including a spatially
periodic potential (rather than the HO potential considered here). Another
interesting generalization is to introduce the \textquotedblleft
nonlinearity management" \cite{Springer}, that would periodically (in time)
switch the system between regions of $\gamma <1$ and $\gamma >1$, where the
two different types of the ground stated are expected, \textit{viz}., the
SVs and MMs, respectively. The periodic switching between them may give rise
to new dynamical regimes.

\end{document}